\documentclass[reprint,showpacs,preprintnumbers,amsmath,amssymb,floatfix,prc]{revtex4-1}
\usepackage{graphicx}
\usepackage{dcolumn}
\usepackage{bm}

\def\beq{\begin{equation}}
\def\eeq{\end{equation}}
\def\be{\begin{eqnarray}}
\def\ee{\end{eqnarray}}

\newcommand{\lsim}{
 \mathrel{\setbox0=\hbox{$<$}\raise0.6ex\copy0\kern-\wd0
 \lower0.65ex\hbox{$\sim$}}}

\newcommand{\gsim}{
 \mathrel{\setbox0=\hbox{$>$}\raise0.6ex\copy0\kern-\wd0
 \lower0.65ex\hbox{$\sim$}}}

\begin{document}
\title{Photoproduction of heavy quarks in 
ultraperipheral pp, pA, and AA collisions at the CERN Large Hadron Collider}
\author{Adeola Adeluyi}
\affiliation{Department of Physics \& Astronomy,
Texas A\&M University-Commerce, Commerce, TX 75428, USA}
\author{Trang Nguyen}
\affiliation{Center for Nuclear Research, Department of Physics \\
Kent State University, Kent, OH 44242, USA}
\date{\today}
\begin{abstract}
Photoproduction of heavy quarks in ultraperipheral collisions 
can help elucidate important features of the physics 
of heavy quarks in Quantum Chromodynamics (QCD). Due to the 
dependence on parton distributions it can also potentially 
offer some constraining ability in the determination of 
nuclear parton distributions.
In the present study we consider  next-to-leading order (NLO) 
photoproduction of heavy quarks in ultraperipheral proton-proton (pp), 
proton-nucleus (pA), and nucleus-nucleus (AA) collisions at  
the CERN Large Hadron Collider (LHC).
Total cross sections and rapidity distributions are considered and  
the influence of nuclear modifications of parton distributions on
these quantities are explored for pA and AA collisions. We find that
photoproduction of heavy quarks in PbPb collisions exhibit significant 
sensitivity to nuclear effects, and in conjunction with
photoproduction in pPb collisions, affords good constraining potential 
for gluon shadowing determination. 

\end{abstract}
\maketitle
\vspace{1cm}
%
%
\section{Introduction}
The study of ultraperipheral relativistic heavy ion collisions 
is useful in exploring several aspects of particle and nuclear physics 
and is an important part of current experimental efforts at the LHC.
Consequently ultraperipheral collisions have been extensively
discussed in the literature (for a small sample of references, see
e.g. \cite{Jackson, Bertulani:1988,Cahn:1990jk,Baur:1990fx,KN99,
Bertulani:1999cq,Goncalves:2001vs,KNV02,Goncalves:2003is,Bertulani:2005ru,
Baltz:2007kq,AyalaFilho:2008zr,Salgado:2011wc}) for different
collision systems and various photon-nucleus and photon-photon processes.

In \cite{Adeluyi:2011rt,Adeluyi:2012ph,Adeluyi:2012ds} we have 
considered photoproduction of $c\bar{c}$ and $b\bar{b}$ in 
ultraperipheral pPb ($\sqrt{s_{_{NN}}}=5$ TeV and 
($\sqrt{s_{_{NN}}}=8.8$ TeV) and PbPb ($\sqrt{s_{_{NN}}}=2.76$ TeV and 
($\sqrt{s_{_{NN}}}=5.5$ TeV) collisions at the LHC. A major
shortcoming was that the results were calculated to leading order (LO) 
in perturbative QCD. In this article we remedy this shortcoming and
present results at next-to-leading order (NLO) accuracy. We also
consider, as an addition to the previous treatments, photoproduction 
of heavy quarks in ultraperipheral pp collisions at
$\sqrt{s_{_{NN}}}=7$ TeV and $\sqrt{s_{_{NN}}}=14$ TeV. As in the
earlier studies an important goal is to investigate quantitatively the
influence of nuclear modifications of parton distributions on 
observables such as cross sections and rapidity distributions and also
the extent of applicability of these observables in helping to 
constrain components of nuclear parton distributions. 

The paper is organized as follows: in Sec.~\ref{phothadupc} 
we discuss photoproduction of heavy quarks and the basic ingredients 
required in our calculations, namely photon flux, photohadron cross
sections, and input nuclear and photon parton distributions.  
In Sec.~\ref{res} we present the results of our calculations. 
Our conclusion is contained in Sec.~\ref{conc}. 

\section{Photon-hadron interactions in ultraperipheral collisions}
\label{phothadupc}
In general the total cross section for the photoproduction of a pair
of heavy quarks ($c\bar{c}$ or $b\bar{b}$) in ultraperipheral
collisions can be expressed as a convolution of the incident 
photon flux, $dN_\gamma/dk$, with the energy-dependent photo-cross
section $\sigma^{\gamma H  \rightarrow Q \overline{Q}}(k)$, with $k$ denoting the 
photon energy and $H$ a proton or nucleus. Thus for pp collisions we can write  
\begin{equation}
\sigma_{pp}^{Q \overline{Q}}=2 \int dk \,\frac{dN_\gamma^{p}(k)}{dk}\, \sigma^{\gamma p
  \rightarrow Q \overline{Q}}(k) \, \, ,
\label{tAAxs}
\end{equation}
with ${dN_\gamma^p}/{dk}$ the photon flux from a proton in
the pp collision system and $\sigma^{\gamma p \rightarrow Q \overline{Q}}(k)$
the energy-dependent cross section for the photoproduction of a 
$Q \overline{Q}$ pair off a proton.   
The factor of $2$ accounts for the source/target symmetry present in symmetric
collisions, since each proton can act simultaneously as a source and
target of photons. A similar expression holds for AA collisions with  
${dN_\gamma^p}/{dk}$ replaced with ${dN_\gamma^Z}/{dk}$, the flux from
one of the participating nuclei and 
$\sigma^{\gamma p \rightarrow Q  \overline{Q}}(k)$ replaced with 
$\sigma^{\gamma A \rightarrow Q \overline{Q}}(k)$, the cross section
for photoproduction of a $Q \overline{Q}$ pair off a nucleus.  
In the case of pA collisions the nucleus 
acts preferentially as the source and the proton as the target, 
leading predominantly to $\gamma$p processes. But there is still 
a non-negligible contribution from $\gamma$A processes in which the 
proton acts as the source of photons and the nucleus as the target. 
Thus expressions for both types of fluxes are required for pA
collisions. We can thus write
\begin{equation}
\sigma_{pA}^{Q \overline{Q}}= \int dk \bigg[\frac{dN_\gamma^Z}{dk}\, \sigma^{\gamma p
  \rightarrow Q \overline{Q}}(k)
+ \frac{dN_\gamma^p}{dk}\, \sigma^{\gamma A
  \rightarrow Q \overline{Q}}(k)\bigg] \, \, ,
\label{tpAxs}
\end{equation}
with ${dN_\gamma^Z}/{dk}$ and ${dN_\gamma^p}/{dk}$ the fluxes from the 
nucleus and proton in the pA collision system and 
$\sigma^{\gamma p \rightarrow Q  \overline{Q}}(k)$ and 
$\sigma^{\gamma A \rightarrow Q  \overline{Q}}(k)$ the photonucleon
and photonuclear cross sections respectively.

Manifestations of nuclear effects can be more transparently observed
when rapidity distributions are considered.
Using the relation $d\sigma/dy = kd\sigma/dk$  
the pair rapidity distribution can be expressed as 
\begin{equation}
\frac{d \sigma^{\gamma H \rightarrow Q \overline{Q}}}{dy} = 
k\, \frac{dN_\gamma(k)}{dk}\, \sigma^{\gamma H
\rightarrow Q \overline{Q}}(k)
\label{Hqqhrap}
\end{equation}
and scales directly with the photon flux $dN_\gamma/dk$.
Thus for pA collisions, with the convention that the proton is
incident from the right and the nucleus from the left, the total 
rapidity distribution is
\begin{eqnarray}
\frac{d \sigma^{pA \rightarrow Q \overline{Q}}}{dy}&=& 
\bigg[k \, \frac{dN_\gamma^Z(k)}{dk}\, \sigma^{\gamma p
\rightarrow Q \overline{Q}}(k)\bigg]_{k=k_l} \nonumber \\
&+&\bigg[k \, \frac{dN_\gamma^p(k)}{dk}\, \sigma^{\gamma A
\rightarrow Q \overline{Q}}(k)\bigg]_{k=k_r} 
\label{pAqqhrap}
\end{eqnarray}
where $k_l$ ($k_l \propto e^{-y}$) and $k_r$ ($k_r \propto e^{y}$) 
simply denote photons from the nucleus and proton respectively.
The first term on the right-hand side
($\gamma$p distribution) peaks at positive rapidities while the 
second term ($\gamma$A distribution) peaks at negative rapidities.
Since both the fluxes and cross sections are different,
the total distribution is manifestly asymmetric, and the $\gamma$p 
term dominates due to the much larger nuclear 
flux ${dN_\gamma^Z}/{dk}$. 

The total rapidity distribution for AA collisions can 
likewise be written as 
\begin{eqnarray}
\frac{d \sigma^{AA \rightarrow Q \overline{Q}}}{dy}&=& 
\bigg[k \, \frac{dN_\gamma^Z(k)}{dk}\, \sigma^{\gamma A
\rightarrow Q \overline{Q}}(k)\bigg]_{k=k_l} \nonumber \\
&+&\bigg[k \, \frac{dN_\gamma^Z(k)}{dk}\, \sigma^{\gamma A
\rightarrow Q \overline{Q}}(k)\bigg]_{k=k_r} 
\label{AAqqhrap}
\end{eqnarray}
with $k_l$ and $k_r$ simply denoting photons from the nucleus 
incident from the left and right respectively. Here the cross sections and
the left/right fluxes are identical; thus the first term on the 
right-hand side of Eq.~(\ref{AAqqhrap}) yields a distribution 
which is the mirror image of the second term. Consequently the 
total distribution is symmetric about midrapidity ($y=0$). 
Similar expression and symmetry attribute hold true for  
pp collisions.
\subsection{Photon flux}
\label{gamflux}
Let us now consider the various photon fluxes occurring in the 
expressions for total cross sections and rapidity distributions.
For a given impact 
parameter ${\bf b}$, the flux of virtual photons with photon energy
$k$, ${d^3N_\gamma(k,{\bf b}) / dkd^2b} $,  is strongly dependent
on the Lorentz factor $\gamma$. 
The photon flux also depends strongly on the  adiabaticity parameter 
 $\zeta=kb/\gamma$ \cite{Bertulani:1988, Cahn:1990jk,Baur:1990fx}:
\begin{equation}
\frac{d^3N_\gamma(k,{\bf b})}{dkd^2b} = 
\frac{Z^2\alpha \zeta^2}{\pi^2kb^2} \left[ K_1^2(\zeta) + \frac{1}{
\gamma^2} K_0^2(\zeta) \right] \, \, ,
\label{dpf}
\end{equation}
which drops off exponentially for $\zeta >1$, above a cutoff 
energy determined  essentially by the size of the nucleus, 
$E_{cutoff} \sim \gamma$MeV$/b$ (fm).

The photon flux from a proton is usually estimated using the dipole
formula for the electric form factor \cite{Drees:1988pp}: 
\begin{eqnarray}
\frac{dN_\gamma^p(k)}{dk} &=& \frac{\alpha}{2\pi k} 
\bigg[1+ \big(1-\frac{2k}{\sqrt{s_{_{NN}}}}\big)^2\bigg]
\nonumber \\
&&\bigg(\ln{D} - \frac{11}{6} + \frac{3}{D} - \frac{3}{2D^2}
+\frac{1}{3D^3}\bigg) \, \, ,
\label{pf}
\end{eqnarray}
where
$D = 1 + [\rm 0.71 \, GeV^2/Q_{\rm min}^2]$ and the minimum momentum 
transferred $Q_{\rm min}^2 = k^2/[\gamma^2(1-2k/\sqrt{s_{_{NN}}})]$. 

In the case of proton-nucleus (pA) collisions, the flux from the
proton is given by Eq.~(\ref{pf})
while the flux due to the nucleus (of charge Z) can be evaluated 
analytically and is given by \cite{Bertulani:1988},
\begin{eqnarray}
\frac{dN_\gamma^Z(k)}{dk} &=& 
{2Z^2 \alpha \over\pi k} \bigg[ \zeta_R^{pA}K_0(\zeta_R^{pA})
K_1(\zeta_R^{pA}) \nonumber \\
&-& {(\zeta_R^{pA})^2\over 2} \big(K_1^2(\zeta_R^{pA})-K_0^2(\zeta_R^{pA})
\big) \bigg] \, \, , 
\label{anaf}
\end{eqnarray}
with reduced adiabaticity parameter, $\zeta_R^{pA}$, given by
$\zeta_R^{pA} = k(R_p + R_A)/\gamma$ and $R_p$ the effective radius 
of the proton. 

For nucleus-nucleus (AA) collisions an analytic approximation 
similar to Eq.~(\ref{anaf}) can be derived. A more accurate 
expression for the flux can be obtained by 
integrating ${d^3N_\gamma(k,{\bf b}) / dkd^2b} $ over impact
parameters with the constraint of no hadronic interactions and 
accounting for the photon polarization. This yields the total 
photon flux ${dN_\gamma^Z(k) / dk} $  given by  \cite{KN99,Baltz:2007kq},
\begin{eqnarray}
\frac{dN_\gamma^Z(k)}{dk}& =&  
2 \pi \int_{2R_A}^{\infty} db \, b 
\int_0^R  {dr \, r \over \pi R_A^2} 
\int_0^{2\pi} d\phi  \nonumber \\
&\times& {d^3N_\gamma(k,b+r\cos \phi)\over dkd^2b} \, \, ,
\label{npf} 
\end{eqnarray}  
with $R_A$ the radius of the nucleus.

\subsection{Photoproduction cross sections}
\label{pphqs}
In general photon interactions with hadrons and nuclei can be 
classified as direct or resolved.
In direct interactions the photon behave as a point-like
particle while in resolved interactions the 
incident photon first fluctuates into a quark-antiquark state (or 
an even more complex partonic configuration consisting of quarks and
gluons) which then subsequently interacts hadronically with the hadron
or nuclear target. The cross section for the photoproduction of a pair
of heavy quarks, 
$\sigma^{\gamma H\to Q\overline{Q}}\, (k)$, is therefore a sum of both the 
direct and resolved contributions:
\begin{equation}
\sigma^{\gamma H \rightarrow Q\overline{Q}}\, (k) = 
\sigma^{\gamma H \rightarrow Q\overline{Q}}_{direct}\, (k) + 
\sigma^{\gamma H \rightarrow Q\overline{Q}}_{resolved}\, (k) .
\end{equation}
Here $H$ stands for a proton or a nucleus ($H \equiv p, A$)
and the total photoproduction cross section is obtained by convolution
of the energy-dependent cross section 
$\sigma^{\gamma H\rightarrow Q\overline{Q}}\, (k)$ 
and the equivalent photon flux, $dN_\gamma(k)/dk$, as given in
Eq.~(\ref{tAAxs}) for symmetric pp (AA) collisions and
Eq.~(\ref{tpAxs}) for asymmetric pA collisions.

Let us consider both contributions in some detail. In view of the 
high energies involved perturbative QCD is applicable, and both 
the direct and resolved contributions are expressible 
as convolution of the cross sections for the relevant partonic 
subprocesses and the corresponding nucleon/nuclear parton distributions. 
Thus the direct contribution can be written as 
\begin{eqnarray}
\sigma^{\gamma H\rightarrow Q\overline{Q}}_{direct}\, (s) =
\sum_{a=q,\bar{q},g}\int dx_a
 \,  \sigma^{\gamma a \rightarrow Q\overline{Q}}(x_as,Q^2) \nonumber \\
 \times f_{a}^{H}(x_a,Q^2)\,\Theta(\zeta_a) \, .
\label{dppcs}
\end{eqnarray}
Here $\sigma^{\gamma a \rightarrow Q\overline{Q}}(x_as,Q^2)$ is the 
parton-level cross section for the photoproduction of a heavy 
quark pair from the interaction of a photon $\gamma$ and a 
parton $a$ with momentum fraction $x_a$. The renormalization 
scale, $\mu_r$, has been set equal to the factorization scale, 
$\mu_f$ i.e. $\mu_r = \mu_f = Q$ and $f_{a}^{H}(x_a, Q^2)$
is the parton distribution of $a$ in $H$ evaluated at $x_a$ and $Q^2$. 
In addition $s=W_{\gamma H}^2$ denotes 
the square of the center-of-mass energy of the photon-nucleus 
(or photon-nucleon) system, $\hat{s}=W_{\gamma a}^2=x_as$ that of the 
photon-parton system,  and $\zeta_a =\hat{s}-4m_{Q}^2$ with $m_Q$ the
mass of the heavy quark (charm or bottom).
The function $\Theta(\zeta_a)$ enforces a minimum (``threshold'') value 
of $x_a$, $x_{a}^{min}$, on the integral given by 
$x_{a}^{min} = 4m_Q^2/s$. Note that the summation over $q$
involves only the light flavors, i.e. $q = u,d,s$.

The resolved contribution is similar in structure to hadroproduction of 
heavy quarks and can be written as
\begin{eqnarray}
\sigma^{\gamma H\rightarrow Q\overline{Q}}_{resolved}\, (s) =
\sum_{a,b=q,\bar{q},g}\int dx_a dx_b \,
 \sigma^{a b \rightarrow Q\overline{Q}}(x_ax_bs,Q^2) \nonumber \\
\times f_{a}^{\gamma}(x_a,Q^2) f_{b}^{H}(x_b,Q^2) \Theta(\zeta_{ab}),
\,\,\,\,\,\,\,  
\label{rppcs}
\end{eqnarray}
where $\sigma^{a b \rightarrow Q\overline{Q}}(x_ax_bs,Q^2)$ is the 
partonic $ab \to Q\bar{Q}$ cross section, 
$f_{a}^{\gamma}(x_a,Q^2)$ ($f_{b}^{H}(x_2,Q^2)$)
is the distribution of parton $a$ ($b$) with momentum fraction $x_a$ 
($x_b$) in a photon ($H$) respectively, $\hat{s}=x_ax_bs$
and $\zeta_{ab} = \hat{s}-4m_{Q}^2$. Similar to the direct contribution, 
the summation over $q$ involves only light quark flavors, i.e. $q =
u,d,s$. 

It is thus clear that the two major ingredients needed for the
determination of the hadronic photoproduction cross sections are 
the partonic (parton-level) cross sections and parton distributions
in nucleon/nuclei and photons. Below we briefly discuss them. 

\subsubsection{Partonic cross sections}
Photon-gluon fusion is the only relevant subprocess for the 
direct contribution at leading order (LO), and the expression for the
cross section can be found in \cite{Gluck:1978bf,JonesWyld,FriStreng78}.
In the case of LO resolved contribution, only the $gg$ and $q\bar{q}$ 
channels are relevant, and the corresponding cross sections can be found in 
\cite{Gluck:1977zm,Combridge:1978kx,Brock:1993sz}.
We have used the next-to-leading order (NLO) results presented in 
\cite{Ellis:1988sb,Nason:1987xz} for both direct and resolved 
contributions. Small-$x$ effects (see \cite{Frixione:1994dv}) which
could potentially be important for charm photoproduction are not 
addressed in the present study. These effects should be somewhat 
lessened in the case of the resolved contribution due to the inherent 
small-$x$ cutoff ($x>0.001$) present in the photon parton
distributions used in this work.

\subsubsection{Parton distributions in nuclei and photons}
We now turn to the consideration of the parton distributions (PDs) relevant to 
the processes considered in the present study. While the direct
contribution to the photoproduction of heavy quarks requires the distributions
of light quarks/antiquarks and gluons in protons and nuclei, the
resolved contribution, in addition, requires these distributions in
photons also. Thus in addition to the usual requirement of nucleon and 
nuclear parton distributions (nPDs), there is also the need for the 
relatively poorly known parton distributions in photons ($\gamma$PDs), thereby 
increasing the level of the theoretical uncertainties in the
calculation of photoproduction of heavy quarks. 

Let us first discuss nuclear parton distributions. It is rather well-known
that the distributions of partons (i.e. quarks and gluons) in 
nuclei are quite different from the distributions in free nucleons,
that is, they are ``modified'' by the complex, many-body effects in the 
nucleus. These nuclear effects are usually 
parametrized in terms of "nuclear modifications" $R_{a}^A(x,Q^2)$ 
which in general depend on the parton specie ($a$), the nucleus ($A$), 
momentum fraction $x$ and scale $Q^2$.  
The nuclear effects can be categorized based on different intervals in $x$.
At small values of $x$ ($x \lesssim 0.04$), we have the phenomenon
generally referred to as shadowing. This is a depletion, in the sense
that in this interval, the distribution of a parton $a$ in the nucleus
is smaller compared to the corresponding distribution in a free 
proton, i.e. $R_a^A < 1$. Antishadowing, which is an
enhancement ($R_a^A > 1$), occurs in the range $0.04 \lesssim x \lesssim
0.3$. Another depletion, the classic EMC effect \cite{Aubert:1983xm}, is
present in the interval $0.3 \lesssim x \lesssim 0.8$, while for 
$x > 0.8$, the Fermi motion region, we have another enhancement. 
It is important to note that although both shadowing and
the EMC effect (antishadowing and Fermi motion) correspond to
depletion (enhancement), the physical principles and mechanisms 
governing these phenomena are quite different. Further details can be
found in \cite{Geesaman:1995yd,Piller:1999wx,Armesto:2006ph,Kolhinen:2005az} 
With the knowledge of $R_{a}^A(x,Q^2)$, nuclear parton distributions 
can be expressed as a convolution of free nucleon parton distributions 
and nuclear modifications, i.e. 
$f_a^A(x,Q^2) = f_a(x,Q^2) \otimes R_{a}^A(x,Q^2)$.

While the determination of quark and antiquark distributions in 
nucleons and nuclei is in general a nontrivial task, that of gluons
is even more problematic. Gluons are electrically neutral, and 
thus their distributions cannot be extracted directly from 
Deeply Inelastic Scattering (DIS) and Drell-Yan (DY) processes 
which account for the major part of the data used in global fits.  
Their distributions are in general inferred from sum
rules and the $Q^2$ evolution of sea quarks distributions. 
The situation is even worse in the nuclear case: the
available data is much less than for nucleons, and there is the added
complication of a mass dependence. It is therefore not unusual for
nuclear gluon distributions from different global fits to differ 
significantly, especially in the magnitude of the various nuclear
effects (shadowing, antishadowing, etc). This is especially obvious
at low $Q^2$ (i.e. around their initial starting scales) since 
evolution to high $Q^2$ tends to lessen the differences.  
Earlier global analyses 
\cite{Eskola:1998df,deFlorian:2003qf,Shad_HKN,Hirai:2007sx} relied
heavily on fixed-target nuclear deep-inelastic scattering
(DIS) and Drell-Yan (DY) lepton-pair production data. 
Incorporation of additional data from inclusive hadron production 
in deuteron-gold collisions and neutrino-iron processes has been 
implemented in \cite{Eskola:2008ca,Eskola:2009uj,Schienbein:2009kk,
Stavreva:2010mw,Kovarik:2010uv,deFlorian:2011fp}.  
The approach in \cite{Frankfurt:2003zd,Frankfurt:2011cs} utilizes the Gribov 
picture of shadowing and data from diffractive processes to generate
nuclear modifications. Despite all these advances the nuclear gluon
distribution is still currently the least constrained aspect of  
nuclear parton distributions, as significant uncertainties still persist at 
both small and large $x$.
    
Three recent nucleon/nuclear parton distributions are utilized in 
the present study. For the proton we use the Martin-Stirling-Thorne-Watts
(MSTW08) parton distributions \cite{Martin:2009iq} which are 
available up to next-to-next-to-leading order (NNLO).
In the nuclear case we use two nuclear modification sets:
the NLO set by  Eskola, Paukunnen, and Salgado (EPS09) 
\cite{Eskola:2009uj} and the NLO set by Frankfurt,
Guzey, and Strikman (FGS10) \cite{Frankfurt:2011cs}. For EPS09 we 
use the central fit while for FGS10 we use the strong 
gluon shadowing model (FGS10\_H).
The distributions from MSTW08 serve two purposes: 
as the free nucleon distributions used in conjunction with
nuclear modifications, and also as a ``special'' nuclear 
distribution in the absence of nuclear effects. The latter case is
particularly useful for highlighting the influence of the various
nuclear effects on observables. 

\begin{figure}[!htb]
\includegraphics[width=\columnwidth]{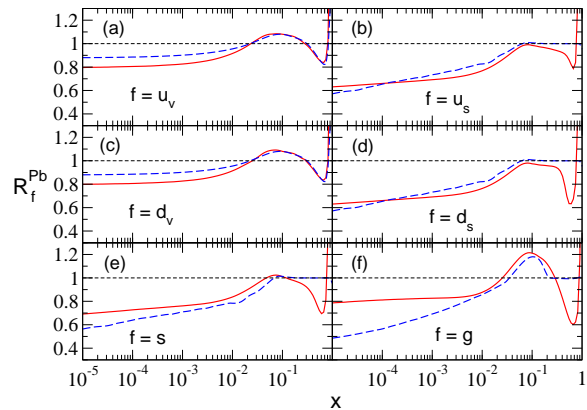}
\caption[...]{\label{fig:RfPb} (Color online) Nuclear modifications of 
light quarks and gluons in Pb, $R_{f}^{Pb}(x,Q^2 = 4m_{c}^2$ GeV$^2$), 
from EPS09 (solid line) and FGS10H (dashed line): (a) valence up
quark $u_{_v}$, (b) up sea quark $u_s$ ($\equiv \bar{u}$), 
(c) valence down quark $d_{_V}$, (d) down sea quark $d_s$ ($\equiv
\bar{d}$), (e) strange quark $s$ ($ \equiv \bar{s}$), and (e) gluon $g$.}
\end{figure}

It is instructive to compare the characteristics of the light 
partons and gluon distributions from both EPS09 and FGS10 sets based 
on the strength of their nuclear modifications. 
In Fig.~\ref{fig:RfPb} we show 
the nuclear modifications for light quarks and gluons in Pb, 
$R_f^Pb(x,Q^2)$ from EPS09 and FGS10 at the factorization scale $Q^2 =
4m_{c}^2$. 
For valence up ($u_{_v}$) and down ($d_{_v}$) quarks, both distributions
exhibit identical antishadowing, EMC, and Fermi motion characteristics.  
Both $u_{_V}$ and $d_{_v}$ are less shadowed in FGS10 than in EPS09 for the 
entire shadowing interval ($\approx 0.9$ versus $0.8$ at very small $x$). 
In the case of $\bar{u}$ and $\bar{d}$, the EMC and Fermi motion
effects which are present in EPS09 are virtually absent in FGS10. The
shadowing components are not very different, with FGS10 being slightly 
more shadowed at very small $x$, and EPS09 more shadowed for $x>10^{-4}$.
In FGS10 strange quarks (antiquarks) are more shadowed but there is an 
absence of EMC and Fermi motion characteristics which are present in 
EPS09. By far the most impressive difference occurs in the shadowing 
of the gluon sector: the gluon shadowing in EPS09 is almost constant 
at small $x$ at $\approx 0.8$ while in FGS10 the shadowing increases 
with decreasing $x$, reaching a value of about $0.49$ 
at $x\approx 10^{-5}$. While EPS09 exhibits both EMC and Fermi motion 
effects these are absent in FGS10. Also the antishadowing component 
is smaller in FGS10 than in EPS09. Thus in the shadowing regime it is 
reasonable to expect that much of the differences in the predictions 
from EPS09 and FGS10 stem from their radically different gluon
characteristics.

Parton distributions in photons ($\gamma$PDs) are derived from experimentally 
determined photon structure function $F_2^{\gamma}(x,Q^2)$, in conjunction 
with appreciable theoretical inputs. These inputs, which are 
necessary in implementing the parametrization of photon parton 
distributions from the structure function, account in part for 
some of the observable differences in the available photon 
parton distribution sets. Another source of differences is in the 
choice and scope of experimental data from which $F_2^{\gamma}$ is 
extracted. At present there is an appreciable number of photon parton 
distribution sets available, both at leading and next-to-leading orders
\cite{Duke:1982bj,Drees:1984cx,Abramowicz:1991yb,Hagiwara:1994ag,
Gluck:1991ee,Gluck:1994tv,Gluck:1991jc,Gordon:1996pm,Gordon:1991tk,
Schuler:1996fc,Schuler:1995fk,Aurenche:1994in,Aurenche:1992sb,Cornet:2002iy,
Cornet:2004ng,Cornet:2004nb,Aurenche:2005da}. It should be noted that  
unlike in the case of a nucleon, there are no valence quarks 
present in the photon; therefore antiquark distributions are the 
same as quark distributions. Furthermore, there are no sum rules 
governing the gluon content; thus the gluon distribution is 
almost totally unconstrained. The gluon distribution contributes to 
$F_2^{\gamma}(x,Q^2)$ majorly through the $\gamma^*g \to q\bar{q}$ 
channel, which has significant numerical support only at small $x$.

\begin{figure}[!htb] 
\includegraphics[width=\columnwidth]{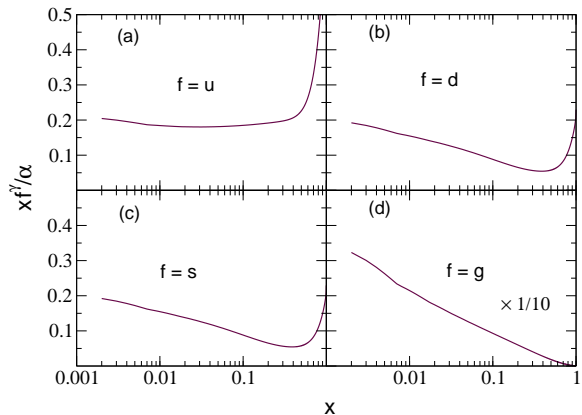}
\caption[...]{\label{photpdfafg04} (Color online) Parton distributions 
in the photon for (a) up, (b) down, (c) strange quarks, and (d) gluons at 
$Q^2 = 4m_{c}^2$ GeV$^2$ from the AFG04 distribution set.  
For better visuality the gluon distribution (d) has been scaled 
down by a factor of $10$.}
\end{figure}

Two recent NLO photon parton distributions are the set by Cornet,
Jankowski, and Krawczyk (CJK04) \cite{Cornet:2004nb} and that by 
Aurenche, Fontannaz, and Guillet (AFG04) \cite{Aurenche:2005da}. 
The CJK04 set uses the DIS renormalization scheme while the MSbar 
scheme is employed in AFG04. We have used the AFG04 set (AFG04\_BF) 
in the present
study for consistency with the renormalization scheme used in  
\cite{Ellis:1988sb,Nason:1987xz} for the NLO cross sections used in 
this work. 
In Fig.~\ref{photpdfafg04} we show the parton 
distributions from the AFG04 NLO set for the light (u,d,s) quarks and gluon.  
The most remarkable characteristic is the dramatic rise of $g^{\gamma}$ 
with decreasing $x$. 
\section{Results}
\label{res}
We now discuss the results of our calculations for the inclusive 
photoproduction of heavy quarks ($c\bar{c}$ and $b\bar{b}$) in 
ultraperipheral proton-proton (pp), proton-lead (pPB), and 
lead-lead (PbPb) collisions at the LHC. The calculations  
are carried out at two center-of-mass (cms) energies for 
each collision system:  
$E_1^{pp}$ ($\sqrt{s_{_{NN}}}=7$ TeV) and $E_2^{pp}$ 
($\sqrt{s_{_{NN}}}=14$ TeV) for pp, 
$E_1^{pPb}$ ($\sqrt{s_{_{NN}}}=5$ TeV) and $E_2^{pPb}$ 
($\sqrt{s_{_{NN}}}=8.8$ TeV) for pPb, and 
$E_1^{PbPb}$ ($\sqrt{s_{_{NN}}}=2.76$ TeV) and $E_2^{PbPb}$ 
($\sqrt{s_{_{NN}}}=5.5$ TeV) for PbPb collisions respectively. 
We take $m_c = 1.4$ GeV and $m_b = 4.75$ GeV for
consistency with the MSTW08 parton distributions.  
The strong coupling constant at scale $Q^{2}$, $\alpha_{s}(Q^{2})$,  
is evaluated to two loops using the evolution code contained in 
the MSTW08 package. For $c\bar{c}$ we set  $Q^{2}=4m_c^2$ while 
for $b\bar{b}$ we take $Q^{2}=m_b^2$.

\subsection{Heavy quarks in pp collisions}
In Table~\ref{ccbbpp} we show the cross sections for the direct 
and resolved components, as well as their sum, for the photoproduction 
of $c\bar{c}$ and $b\bar{b}$ in ultraperipheral pp collisions at 
the two specified cms energies. 
\begin{table}[!htb]
\caption{\label{ccbbpp} Direct, resolved, and total cross sections 
for photoproduction of $c\bar{c}$ and $b\bar{b}$ in ultraperipheral 
pp collisions at the LHC. All cross sections are in nanobarns (nb).}
\begin{tabular}[c]{|l | c| c| c| c|}
\hline
            & $\sqrt{s_{_{NN}}}$ (TeV)  & Direct  & Resolved  & Total \\
\hline
$c\bar{c}$  & $7$              & 2050.9  & 677.1   & 2728.0 \\ 
            & $14$             & 2837.6  & 1124.7 & 3962.3 \\
\hline
$b\bar{b}$  & $7$              & 31.2   & 20.1    & 51.3 \\
            & $14$             & 52.8   & 44.0    & 96.8 \\ 
\hline
\end{tabular} 
\end{table}

Some interesting features can be deduced from Table~\ref{ccbbpp}. Thus 
at $\sqrt{s_{_{NN}}} = 7$ TeV the resolved contribution is about 
$24.8\%$ of the total cross section for $c\bar{c}$ and $39.1\%$ for 
$b\bar{b}$, while at $\sqrt{s_{_{NN}}} = 14$ TeV the resolved
component accounts for $28.4\%$ and $45.5\%$ respectively for 
$c\bar{c}$ and $b\bar{b}$ production. It is thus evident that 
the resolved contribution, being a significant fraction of the 
total cross section, is important for $c\bar{c}$, and more especially 
so for $b\bar{b}$ production, with this importance increasing with 
increasing energy.
 
\begin{figure}[!htb] 
\includegraphics[width=\columnwidth]{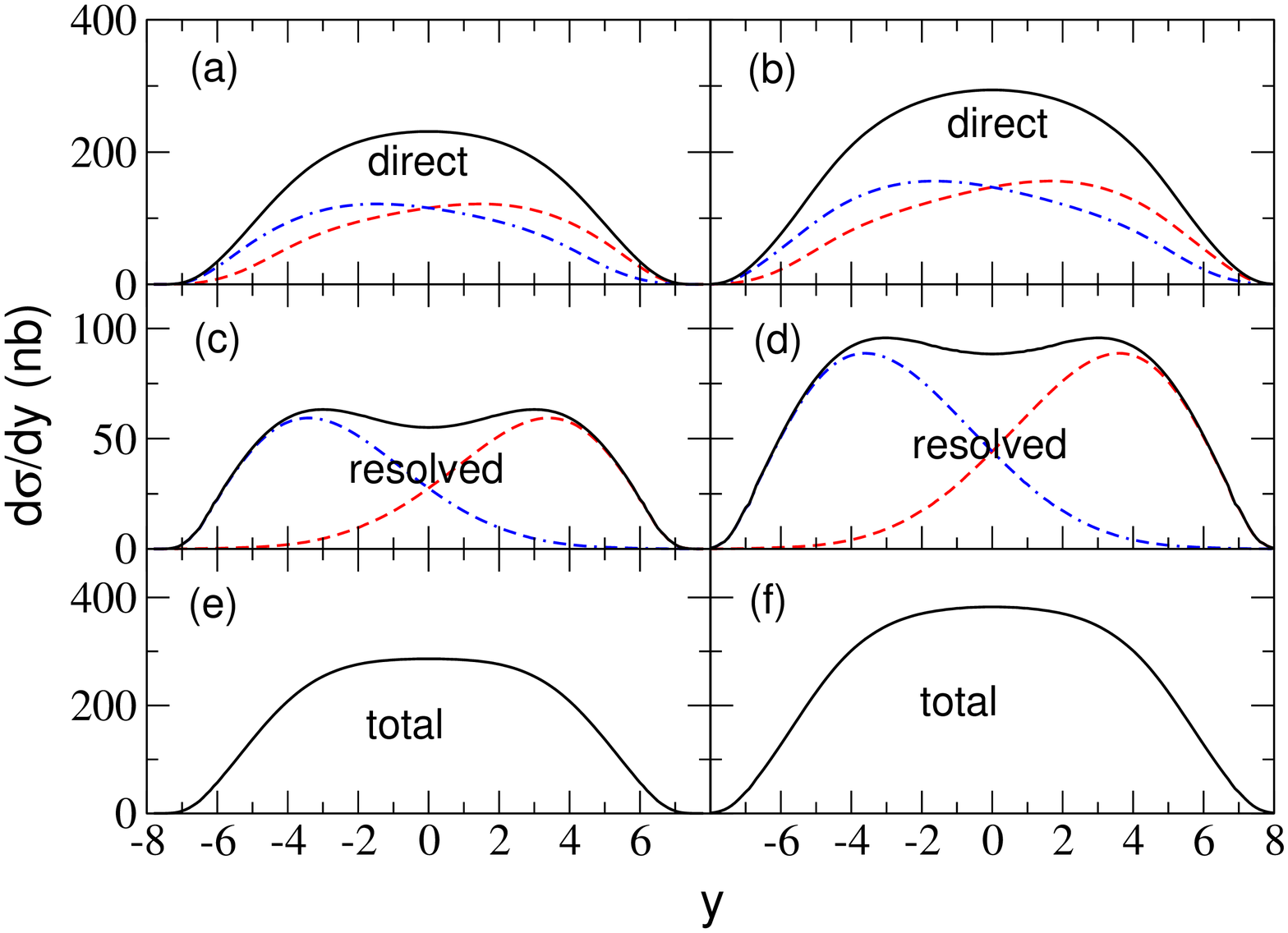}
\caption[...]{\label{ppcc} (Color online) Rapidity distributions of 
$c\bar{c}$ photoproduction in ultraperipheral pp collisions at the LHC.  
Solid lines in (a), (c), and (e)  depict the direct, resolved, and
total distributions at $\sqrt{s_{_{NN}}}=7$ TeV while solid lines 
in (b), (d), and (f) are the equivalent distributions at  
$\sqrt{s_{_{NN}}}=14$ TeV. In the two upper rows dashed and dash-dotted
lines denote contributions from photons incident from the left and 
right respectively.}
\end{figure}

\begin{figure}[!htb] 
\includegraphics[width=\columnwidth]{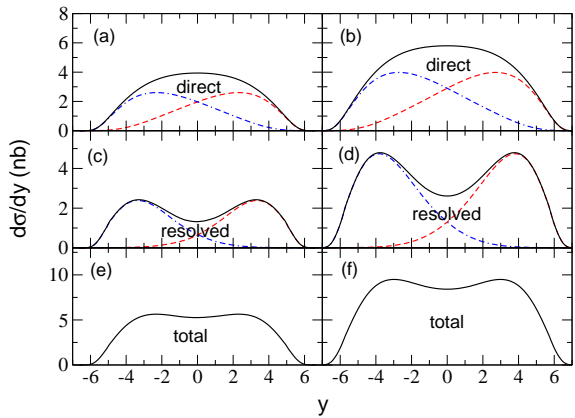}
\caption[...]{\label{fig:ppbb} (Color online) Rapidity distributions of 
$b\bar{b}$ photoproduction in ultraperipheral pp collisions at the LHC.  
Solid lines in (a), (c), and (e)  depict the direct, resolved, and
total distributions at $\sqrt{s_{_{NN}}}=7$ TeV while solid lines 
in (b), (d), and (f) are the equivalent distributions at  
$\sqrt{s_{_{NN}}}=14$ TeV. In the two upper rows dashed and dash-dotted
lines denote contributions from photons incident from the left and 
right respectively.} 
\end{figure}

These features are apparent in the rapidity distributions for 
$c\bar{c}$ as shown in Fig.~\ref{ppcc}, and for $b\bar{b}$ as 
displayed in Fig.~\ref{fig:ppbb}. These distributions are manifestly 
symmetric about midrapidity ($y = 0$) due to the source/target 
%
symmetry present in pp collisions.

It is instructive to explore how the cross sections scale with
energy. To this end we consider the ratio 
of the total cross section at $\sqrt{s_{_{NN}}} = 14$ TeV to that  
at $\sqrt{s_{_{NN}}} = 7$ TeV for both $c\bar{c}$ (denoted by 
$R_{c\bar{c}}$) and $b\bar{b}$ (denoted by $R_{b\bar{b}}$) production.
From Table~\ref{ccbbpp} we can see that $R_{c\bar{c}} = 1.452$ and 
$R_{b\bar{b}} = 1.887$ respectively. Thus
for $c\bar{c}$ the total cross section at $\sqrt{s_{_{NN}}} = 14$
TeV is approximately $1.5$ times larger than at $\sqrt{s_{_{NN}}} = 7$
TeV, and approximately $1.9$ times larger in the case of 
$b\bar{b}$ production. This implies that the cross section for 
$b\bar{b}$ production grows more rapidly with increasing energy.
Componentwise the ratio of direct cross
sections is $\approx 1.4$ for $c\bar{c}$ and $\approx 1.7$ for
$b\bar{b}$ while for resolved it is $\approx 1.7$ and $\approx 2.2$ 
respectively.
\subsection{Heavy quarks in pPb collisions}
Photoproduction in ultraperipheral pPb collisions arise from two 
distinct processes: $\gamma$p interactions in which the Pb nucleus 
is the source of the photon and the proton acts as the target, and 
$\gamma$Pb interactions in which the proton is the source of photons 
and the Pb nucleus is the target. Due to the large disparity in the 
magnitude of the photon fluxes from the proton and the nucleus (Pb) 
the $\gamma$p contribution dominates, and a common
practice is to neglect the photonuclear ($\gamma$Pb) contribution. 
We retain the photonuclear contribution in this study and test the 
severity of the approximation made in neglecting it. 

In Table~\ref{ccbbpA} we present the cross sections for the direct 
and resolved components of both $\gamma$p and $\gamma$Pb contributions for 
$c\bar{c}$ and $b\bar{b}$ production in ultraperipheral 
pPb collisions at the LHC. Total cross sections for both 
$c\bar{c}$ and $b\bar{b}$ are presented in Table~\ref{tccbbpA}. 
\begin{table}[!htb]
\caption{\label{ccbbpA} Direct and resolved cross sections for
 photoproduction of $c\bar{c}$ and $b\bar{b}$ in ultraperipheral pPb 
collisions at the LHC. All cross sections are in microbarns ($\mu$b).}
\begin{tabular}[c]{|l|c|c| c c| c c| }
\hline
   & & &$\sqrt{s_{_{NN}}}$ & $= 5$ TeV & $\sqrt{s_{_{NN}}}$ & $= 8.8$ TeV  \\
\cline{4-7}
   &  &PDF & Direct  & Resolved  &  Direct  & Resolved \\
\hline
 & $\gamma$p & MSTW08 & 2474.6 & 462.2 & 3563.5 & 823.8  \\ 
\cline{2-7}
$c\bar{c}$ & & MSTW08 & 179.6  & 53.8  & 238.2  & 84.0   \\
& $\gamma$Pb & EPS09  & 157.9  & 48.7  & 207.2  & 74.4   \\
&           & FGS10  & 131.9  & 46.5  & 169.5  & 70.5   \\
\hline
& $\gamma$p  & MSTW08 & 21.4 & 6.4       & 38.0     & 15.1  \\ 
\cline{2-7}
$b\bar{b}$  &  & MSTW08 & 2.5  & 1.4       & 3.9      & 2.7   \\
& $\gamma$Pb & EPS09  & 2.3  & 1.4       & 3.6      & 2.6   \\
&            & FGS10  & 2.1  & 1.3       & 3.2      & 2.6   \\
\hline
\end{tabular} 
\end{table}

Let us first consider $c\bar{c}$ production. At $\sqrt{s_{_{NN}}}= 5$ TeV 
the $\gamma$Pb contribution to the total cross section is
approximately $7.4\%$ for MSTW08 (no nuclear modifications), 
$6.6\%$ for EPS09 (moderate gluon shadowing), and $5.7\%$ for FGS10 (strong 
gluon shadowing) respectively. The resolved component accounts for 
$16.3\%$ of total cross sections, and is independent of the choice of 
parton distribution set. Overall the sensitivity to nuclear
modifications (dominantly shadowing) is very small: the
no-modification (MSTW08) total cross section is
reduced by approximately $0.9\%$ and $1.7\%$ by the modifications in 
EPS09 and FGS10 respectively. 

The trend is similar at $\sqrt{s_{_{NN}}}= 8.8$ TeV. Here the 
$\gamma$Pb contribution to the total cross section is
approximately $6.8\%$ for MSTW08, $6\%$ for EPS09, and $5.2\%$ for 
FGS10 respectively. The resolved component yields approximately
$19.3\%$ of total cross sections, and is also independent of the choice of 
parton distribution set. Identically the sensitivity to nuclear
modifications remains the same as at $\sqrt{s_{_{NN}}}= 5$ TeV. 
\begin{figure}[!htb] 
\includegraphics[width=\columnwidth]{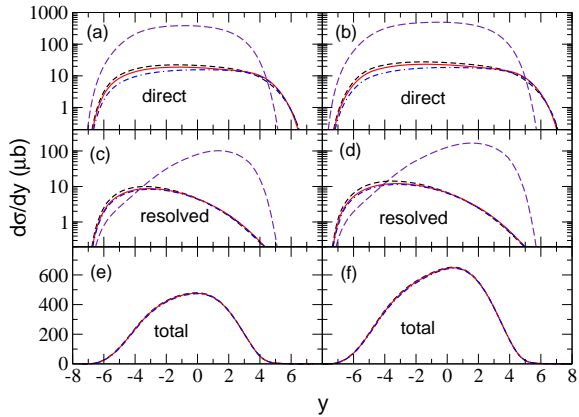}
\caption[...]{\label{fig:pAcc} (Color online) Rapidity distributions 
of $c\bar{c}$ photoproduction in ultraperipheral pPb collisions at the LHC. 
In (a), (c), and (e) we show the direct,
resolved, and total distributions at $\sqrt{s_{_{NN}}}=5$ TeV while  
(b), (d), and (f) display the equivalent distributions 
at $\sqrt{s_{_{NN}}}=8.8$ TeV. In the first and second rows long-dashed 
line depicts the $\gamma$p component while short-dashed (MSTW08), solid 
(EPS09), and dash-dotted (FGS10) lines correspond to $\gamma$Pb 
component with no shadowing, moderate shadowing, and strong
shadowing respectively. In the third row short-dashed, solid, and 
dash-dotted lines depict total distributions using MSTW08, EPS09, 
and FGS10 parton distributions respectively.}
\end{figure}

These features are reflected in Fig.~\ref{fig:pAcc} where we show 
the component as well as total rapidity distributions at 
$\sqrt{s_{_{NN}}}=5$ TeV and $\sqrt{s_{_{NN}}}=8.8$ TeV respectively.
In accordance with our convention the $\gamma$p contribution peaks at 
positive rapidities while the $\gamma$Pb distributions peak at
negative rapidities, and the asymmetric nature of the total
distributions is clearly exhibited. 
Due to the smallness of the $\gamma$Pb contribution, the effects of nuclear 
modifications are totally negligible in the total rapidity
distributions as the distributions from the three different parton 
distribution sets practically overlap.

We now discuss the case for $b\bar{b}$ production. In line with 
the treatment of $c\bar{c}$ production, we determine the relative importance 
of $\gamma$Pb and resolved contributions to the total $b\bar{b}$
production cross sections. At $\sqrt{s_{_{NN}}}= 5$ TeV 
the $\gamma$Pb contribution to the total cross section is
approximately $12.1\%$ for MSTW08, $11.6\%$ for EPS09, and 
$11\%$ for FGS10 respectively. It is immediately apparent that 
at this energy the $\gamma$Pb contributions are more significant than 
their counterparts in $c\bar{c}$ production, and also less sensitive
to nuclear effects. The resolved component is responsible for 
$24.6\%$ of total cross sections and is independent of the choice of 
parton distribution set. The sensitivity to nuclear
modifications is smaller than in $c\bar{c}$ production: the
no-modification (MSTW08) total cross section is
reduced by approximately $0.05\%$ and $1.2\%$ respectively by 
the modifications in  EPS09 and FGS10 parton distribution sets. 

\begin{table}[!htb]
\caption{\label{tccbbpA} Total cross sections for photoproduction of 
$c\bar{c}$ and $b\bar{b}$ in ultraperipheral pPb collisions at
$E_1^{pPb}$ ($\sqrt{s_{_{NN}}}=5$ TeV) and 
$E_2^{pPb}$ ($\sqrt{s_{_{NN}}}=8.8$ TeV). 
All cross sections are in microbarns ($\mu$b).}
\begin{tabular}[c]{|l| c c c| c c c| }
\hline
      & &  $c\bar{c}$ & & & $b\bar{b}$ & \\
\cline{2-7}
PDF&$E_1^{pPb}$&$E_2^{pPb}$&$R_{c\bar{c}}$&$E_1^{pPb}$ & $E_2^{pPb}$&$R_{b\bar{b}}$ \\
\hline
 MSTW08 & 3170.3 & 4709.5&1.486& 31.6 & 59.8 &1.894 \\ 
 EPS09  & 3143.4 & 4668.9&1.485& 31.4 & 59.4 &1.891 \\
 FGS10  & 3115.3 & 4627.3&1.485& 31.2 & 59.0 &1.892  \\
\hline
\end{tabular} 
\end{table}
Similar features are present at $\sqrt{s_{_{NN}}}=8.8$ TeV.
The $\gamma$Pb contribution to the total cross section here is
approximately $11.1\%$ for MSTW08, $10.4\%$ for EPS09, and $9.8\%$ for 
FGS10 respectively. Again these values are relatively more significant
than those obtained in $c\bar{c}$ production. The resolved component 
accounts for approximately $30\%$ of total cross sections, 
and is again independent of the choice of 
parton distribution set. The sensitivity to nuclear
modifications are similar to those obtained at $\sqrt{s_{_{NN}}}=5$
TeV: $0.07\%$ for EPS09 and $1.4\%$ for FGS10 respectively.
\begin{figure}[!htb] 
\includegraphics[width=\columnwidth]{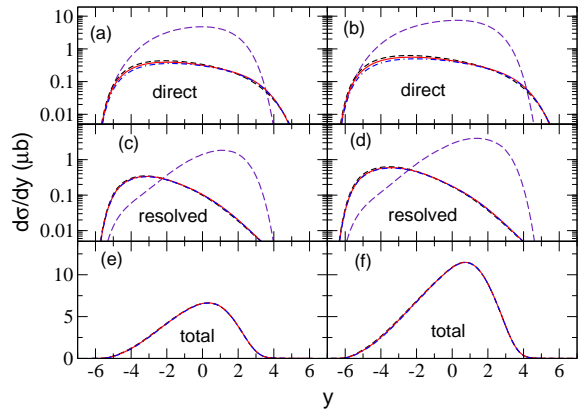}
\caption[...]{\label{fig:pAbb} (Color online) Rapidity distributions 
of $b\bar{b}$ photoproduction in ultraperipheral pPb collisions at the LHC. 
In (a), (c), and (e) we show the direct,
resolved, and total distributions at $\sqrt{s_{_{NN}}}=5$ TeV while  
(b), (d), and (f) display the equivalent distributions 
at $\sqrt{s_{_{NN}}}=8.8$ TeV. In the first and second rows long-dashed 
line depicts the $\gamma$p component while short-dashed (MSTW08), solid 
(EPS09), and dash-dotted (FGS10) lines correspond to $\gamma$Pb 
component with no shadowing, moderate shadowing, and strong
shadowing respectively. In the third row short-dashed, solid, and 
dash-dotted lines depict total distributions using MSTW08, EPS09, 
and FGS10 parton distributions respectively.}
\end{figure}         

In Fig.~\ref{fig:pAbb} we show the component as well as 
total rapidity distributions for $b\bar{b}$ production at 
$\sqrt{s_{_{NN}}}=5$ TeV and $\sqrt{s_{_{NN}}}=8.8$ TeV respectively.
These distributions faithfully manifest the features described 
above. As in $c\bar{c}$ production
the $\gamma$p contribution peaks at positive rapidities while 
the $\gamma$Pb distributions peak at negative rapidities, the 
total distributions being thus asymmetric due to the dominance 
of the $\gamma$p contribution. The effects of nuclear 
modifications on total rapidity distributions are totally negligible, 
as the distributions from the three different parton 
distribution sets practically overlap. This is due to the fact that
even though the $\gamma$Pb contributions are relatively larger, they 
are also almost totally insensitive to nuclear effects.
 
Let us now discuss the magnitude of the change in cross section with 
energy. In Table~\ref{tccbbpA} we present the ratio 
of the total cross section at $\sqrt{s_{_{NN}}} = 8.8$ TeV to that  
at $\sqrt{s_{_{NN}}} = 5$ TeV for $c\bar{c}$, $R_{c\bar{c}}$, 
and for $b\bar{b}$, $R_{b\bar{b}}$. As can be seen
from Table~\ref{tccbbpA} $R_{c\bar{c}} \approx 1.5$ and 
$R_{b\bar{b}} \approx 1.9$ for all three parton distribution sets.
These values are practically the same as obtained for pp collisions.
\subsection{Heavy quarks in PbPb collisions}

We now discuss production of $c\bar{c}$ and $b\bar{b}$ in 
ultraperipheral PbPb collisions at energies $\sqrt{s_{_{NN}}}=2.76$
TeV and $\sqrt{s_{_{NN}}}=5.5$ TeV. In Table~\ref{ccbbAA} we show the 
cross sections for both direct and resolved components while
total cross sections are presented in Table~\ref{tccbbAA}
\begin{table}[!htb]
\caption{\label{ccbbAA} Direct and resolved cross sections for 
photoproduction of $c\bar{c}$ and $b\bar{b}$ in ultraperipheral PbPb 
collisions at the LHC. All cross sections are in millibarns (mb).}
\begin{tabular}[c]{|l|c| c c| c c| }
\hline
     & & $\sqrt{s_{_{NN}}}$ &$ =2.76$ TeV & $\sqrt{s_{_{NN}}}$ & $= 5.5$ TeV  \\
\cline{3-6}
& PDF    & Direct & Resolved  &  Direct  & Resolved \\
\hline
          & MSTW08 & 552.6  & 69.5       & 926.2     & 161.5  \\ 
$c\bar{c}$& EPS09  & 517.6  & 69.6       & 845.9     & 156.9   \\
          & FGS10  & 475.7  & 67.6       & 755.4     & 152.0   \\
\hline
          & MSTW08 & 3.24  & 0.60       & 7.52     & 2.06  \\ 
$b\bar{b}$& EPS09  & 3.30  & 0.64       & 7,40     & 2.14   \\
          & FGS10  & 3.18  & 0.62       & 7.10     & 2.10   \\
\hline
\end{tabular} 
\end{table}

Due to both participants being nuclei, nuclear effects are more 
pronounced in PbPb collisions  than in pPb collisions. Thus the 
resolved components, which in pPb collisions are quite independent 
of the choice of parton distribution set, and so by implication 
insensitive to nuclear effects, now acquire some sensitivity in 
PbPb collisions. Total cross sections and rapidity distributions 
also exhibit enhanced sensitivity.

As usual let us first consider $c\bar{c}$ production.
At $\sqrt{s_{_{NN}}}=2.76$ TeV the resolved components are $11.2\%$, 
$11.9\%$, and $12.5\%$ respectively for MSTW08, EPS09, and FGS10 
distribution sets. In addition, the MSTW08 total cross section is 
reduced by $5.6\%$ and $12.7\%$ respectively by the nuclear 
modifications in EPS09 and FGS10 sets.
 
The same trend is observed at $\sqrt{s_{_{NN}}}=5.5$ TeV. Here the 
resolved components are $14.9\%$ (MSTW08), $15.6\%$ (EPS09), and 
$16.8\%$ (FGS10) respectively. The influence of shadowing is  
more appreciable; the no-modification MSTW08 total cross section is 
reduced by $7.8\%$ (EPS09) and $16.6\%$ (FGS10) respectively. 

\begin{figure}[!htb] 
\includegraphics[width=\columnwidth]{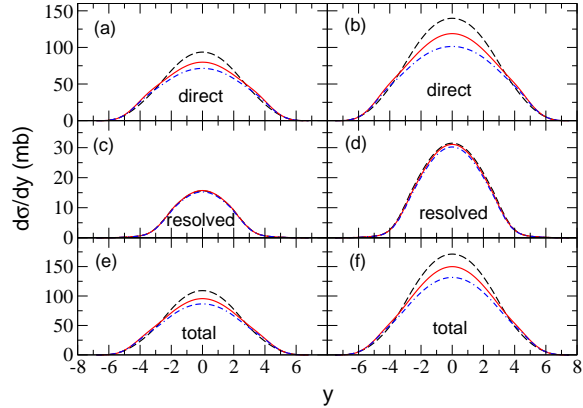}
\caption[...]{\label{AAcc} (Color online) Rapidity distributions of 
$c\bar{c}$ photoproduction in ultraperipheral PbPb collisions at the LHC. 
In (a), (c), and (e) we show the 
direct, resolved, and total distributions respectively 
at $\sqrt{s_{_{NN}}}=2.76$ TeV 
while (b), (d), and (f) denote the equivalent distributions  
at $\sqrt{s_{_{NN}}}=5.5$ TeV. In all cases long-dashed (MSTW08), solid 
(EPS09), and dash-dotted (FGS10) lines correspond to no shadowing, 
moderate shadowing, and strong shadowing respectively.} 
\end{figure}

These features are manifested in the rapidity distributions 
shown in Fig.~\ref{AAcc}. The distributions are symmetric about 
midrapidity ($y = 0$) as expected for symmetric collision systems.
Also the sensitivity to nuclear modifications is more transparent 
here than in total cross sections. 
%
 
Shadowing is the dominant nuclear effect for $-2.5 < y < 2.5$ 
at $\sqrt{s_{_{NN}}}=2.76$ TeV  and for $-3 < y < 3$ at 
$\sqrt{s_{_{NN}}}=5.5$ TeV. The rapidity
distributions in these intervals reproduce the observed trend of 
gluon shadowing strength exhibited in  Fig.~\ref{fig:RfPb}.
MSTW08 with its zero gluon shadowing gives the largest rapidity
distribution while FGS10H, with its strong gluon shadowing, gives the
smallest. Due to strong flux suppression, shadowing is most markedly 
apparent for the rapidity range $-1.5 \lesssim y \lesssim 1.5$ for 
$\sqrt{s_{_{NN}}}=2.76$ TeV and $-2 \lesssim y \lesssim 2$ for 
$\sqrt{s_{_{NN}}}=5.5$ TeV. These intervals are therefore particularly
suited for constraining purposes and also for discriminating among 
different gluon shadowing scenarios.

At $\sqrt{s_{_{NN}}}=2.76$ TeV the rapidity intervals  $2.5 < y < 5$ 
corresponds to $x_{min}$ in the antishadowing region (deep shadowing) 
for right (left) incident photons and vice versa for $-5 < y < -2.5$. 
Due to the photon flux suppression 
in the deep shadowing region, the rapidity distributions are 
sensitive mainly to  antishadowing in addition to both EMC effect and 
Fermi motion. Since both EPS09 and FGS10H have substantial
antishadowing, their rapidity distributions reflect this, 
being slightly larger than those from MSTW08. The
discriminatory power here is not as appreciable as in the shadowing
case though, due largely to the smallness of the distributions. This
is also the case for the rapidity intervals $3 < y < 6$ and 
$-6 < y < -3$ at $\sqrt{s_{_{NN}}}=2.76$ TeV.
\begin{table}[!htb] 
\caption{\label{tccbbAA} Total cross sections for photoproduction of 
$c\bar{c}$ and $b\bar{b}$ in ultraperipheral PbPb collisions at 
$E_1^{PbPb}$ ($\sqrt{s_{_{NN}}}=2.76$ TeV) and 
$E_2^{PbPb}$ ($\sqrt{s_{_{NN}}}=5.5$ TeV). 
All cross sections are in millibarns (mb).}
\begin{tabular}[c]{|l| c c c| c c c| }
\hline
      & &  $c\bar{c}$ & & & $b\bar{b}$ & \\
\cline{2-7}
PDF &$E_1^{PbPb}$&$E_2^{PbPb}$&$R_{c\bar{c}}$&
      $E_1^{PbPb}$&$E_2^{PbPb}$&$R_{b\bar{b}}$\\
\hline
 MSTW08 & 622.0  & 1087.7 &1.75 & 3.9  & 9.6 &2.49 \\ 
 EPS09  & 587.2  & 1002.8 &1.71 & 3.9  & 9.5 &2.42 \\
 FGS10  & 543.3  & 907.3  &1.67 & 3.8  & 9.2 &2.41 \\
\hline
\end{tabular} 
\end{table}

Let us now consider the corresponding case of $b\bar{b}$ production in 
ultraperipheral PbPb collisions. The sensitivity to nuclear
modifications, while more appreciable here than in pPb collisions, 
is significantly less than for the equivalent $c\bar{c}$ production. 
At $\sqrt{s_{_{NN}}}=2.76$ TeV the resolved components are $15.7\%$, 
$16.2\%$, and $16.3\%$ respectively for MSTW08, EPS09, and FGS10 
distribution sets. Thus to a good approximation the resolved component 
can be taken as $16\%$ for all three parton distribution sets.
It is noteworthy that unlike what obtains in $c\bar{c}$ production 
at this energy, both the direct and resolved components from EPS09 are 
the largest. In addition the resolved component from FGS10 is also
larger than that of MSTW08. This is due to the increasingly important 
role of antishadowing at this energy. Thus the EPS09 
total cross section is $2.2\%$ larger than that of MSTW08.
Due to the stronger shadowing and smaller antishadowing in FGS10, 
the total cross section is approximately $1.1\%$ smaller.  

At $\sqrt{s_{_{NN}}}=5.5$ TeV the resolved components are respectively 
$21.5\%$ for MSTW08, $22.4\%$ for EPS09, and $22.9\%$ for FGS10. 
While the influence of shadowing can be seen in the direct components, 
both EPS09 and FGS10 yield larger resolved components, again due to 
antishadowing. Thus the MSTW08 total cross section is reduced by 
$0.6\%$ and $4.4\%$ by EPS09 and FGS10 respectively.

\begin{figure}[!htb] 
\includegraphics[width=\columnwidth]{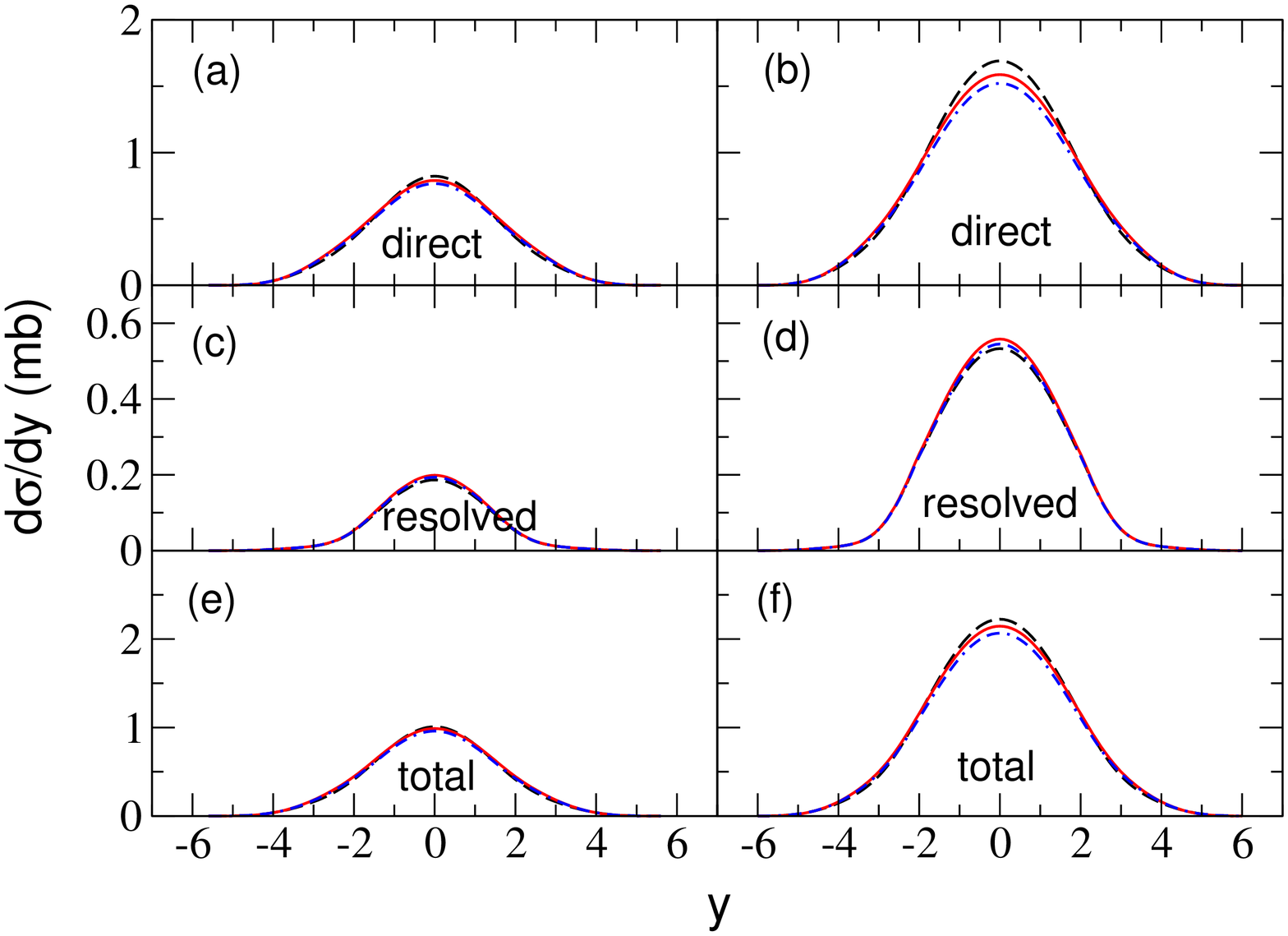}
\caption[...]{\label{fig:AAbb} (Color online) Rapidity distributions of 
$b\bar{b}$ photoproduction in ultraperipheral PbPb collisions at the LHC. 
In (a), (c), and (e) we show the 
direct, resolved, and total distributions respectively 
at $\sqrt{s_{_{NN}}}=2.76$ TeV 
while (b), (d), and (f) denote the equivalent distributions  
at $\sqrt{s_{_{NN}}}=5.5$ TeV. In all cases long-dashed (MSTW08), solid 
(EPS09), and dash-dotted (FGS10) lines correspond to no shadowing, 
moderate shadowing, and strong shadowing respectively.}
\end{figure}

In Fig.~\ref{fig:AAbb} we show the corresponding rapidity distributions. 
Considering the left-hand panels ($\sqrt{s_{_{NN}}}=2.76$ TeV), it is 
apparent that the influence of shadowing present in the direct
component in the rapidity 
interval $-1 < y < 1$ is almost totally negated by the antishadowing 
dominant in the same interval in the resolved component, with the
result that in this interval total rapidity distributions have very 
little sensitivity to nuclear effects. For $-3 < y < -1$ and 
$1 < y < 3$ the slight manifestation of antishadowing seen in the 
direct component persists in the total distributions.

These features are also present in the distributions at 
$\sqrt{s_{_{NN}}}=5.5$ TeV. Here also the clear influence 
of shadowing in the rapidity interval $-2 < y < 2$ in the 
direct component is reduced by the antishadowing present in the 
resolved component in the same interval. Thus the total 
rapidity distribution exhibit reduced sensitivity to nuclear 
shadowing in this interval due to this destructive interference.
Nevertheless the influence of shadowing is still apparent 
especially in the rapidity window $-1 < y < 1$. Thus 
this interval presents the best sensitivity to shadowing effects 
in $b\bar{b}$ production. As at the lower energy, there is a slight 
manifestation of antishadowing in the intervals $-4 < y < -2$ and 
$2 < y < 4$.

Let us now consider the magnitude of the change in cross section with 
energy. In Table~\ref{tccbbAA} we present the ratio 
of the total cross section at $\sqrt{s_{_{NN}}} = 5.5$ TeV to that  
at $\sqrt{s_{_{NN}}} = 2.76$ TeV for $c\bar{c}$, $R_{c\bar{c}}$, 
and for $b\bar{b}$, $R_{b\bar{b}}$. As is readily apparent from the 
Table, both $R_{c\bar{c}}$ and $R_{b\bar{b}}$ vary slightly with the 
choice of parton distribution set. Despite this variation, to a 
good approximation $R_{c\bar{c}} \approx 1.7$ and 
$R_{b\bar{b}} \approx 2.4$ for all three parton distribution sets.
These values are larger, especially for  $R_{b\bar{b}}$ than obtained 
for both pp and pPb collisions.   
\subsection{Theoretical errors and ratio of cross sections} 
From a consideration of the general structure of the quantities 
(cross sections and rapidity distributions) considered in this study, 
three major sources of theoretical errors can be readily identified: 
\begin{itemize}
\item accuracy of the relevant expressions for photon flux, 
\item higher-order corrections, 
\item uncertainties in parton distributions. 
\end{itemize}  
Due to technical reasons we have not attempted to estimate quantitatively  
the uncertainties on calculated quantities in the present study.  
A pragmatic approach to facilitate shadowing determination is to 
compare photoproduction in proton-nucleus and nucleus-nucleus
collisions, where many theoretical uncertainties and systematic errors 
cancel (see for instance \cite{KNV02,Salgado:2011wc}). Here we explore 
in a rather simplistic and nonrigorous manner such an approach for 
shadowing effects in  $c\bar{c}$ and $b\bar{b}$ photoproduction. We 
are primarily interested in relative effects, and as such issues 
like exact normalization and flux parallelism have been ignored.
\begin{figure}[!htb] 
\includegraphics[width=\columnwidth]{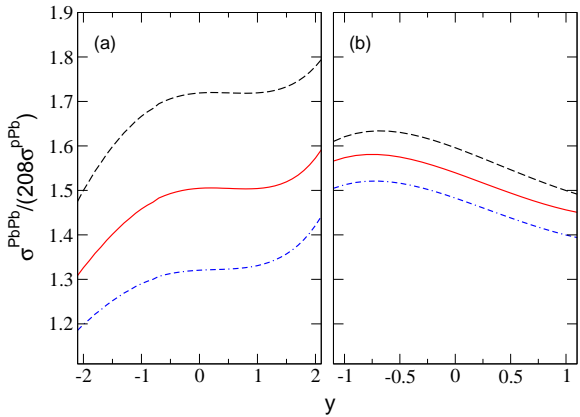}
\caption[...]{\label{rathq} (Color online) Ratio of photoproduction
cross sections, $\sigma^{PbPb}/(208\sigma^{pPb})$, in the shadowing 
region for (a) $c\bar{c}$ and (b) $b\bar{b}$. The PbPb collisions are 
at $\sqrt{s_{_{NN}}}=5.5$ TeV while the pPb collisions are at 
$\sqrt{s_{_{NN}}}=5$ TeV. In both panels long-dashed (MSTW08), solid 
(EPS09), and dash-dotted (FGS10) lines correspond to no shadowing, 
moderate shadowing, and strong shadowing respectively.}  
\end{figure}

In Fig.~\ref{rathq} we show the ratio
$\sigma^{PbPb}/(208\sigma^{pPb})$ as a function of rapidity for both 
$c\bar{c}$ and $b\bar{b}$ in their respective shadowing regions. 
In order to have nearly identical Lorentz factor $\gamma$
the PbPb cross sections, $\sigma^{PbPb}$, have been calculated at 
$\sqrt{s_{_{NN}}}=5.5$ TeV and the pPb cross sections, $\sigma^{pPb}$, 
at $\sqrt{s_{_{NN}}}=5$ TeV. It should be noted that in our
calculations equal rapidity corresponds to equal photon energy.

Let us compare the shadowing effects from these ratios at a 
specific rapidity, say at $y=0$. Thus for $c\bar{c}$ the MSTW08 
value is reduced by approximately $12.5\%$ and $23.2\%$ respectively 
by the shadowing in EPS09 and FGS10. These reductions are precisely 
what one obtains from the rapidity distributions of $c\bar{c}$ in 
PbPb collisions at $\sqrt{s_{_{NN}}}=5.5$ TeV. This is also true for 
$b\bar{b}$ where the ratio reduction values $3.5\%$ (EPS09) and 
$7.1\%$ (FGS10) are the same as from the rapidity distributions. 
It thus seems feasible, at least in this simplistic case, that 
relative shadowing effects are transmitted without appreciable loss.
Of course, in order to ascertain to what degree higher order effects 
are cancelled in these ratios, one should carry out the analogous 
calculations at leading order. Further work along this line is in progress.

\section{Conclusions}
\label{conc}
In the present study we have considered next-to-leading order (NLO) 
photoproduction of $c\bar{c}$ and $b\bar{b}$ in ultraperipheral 
proton-proton (pp), proton-lead (pPb) and lead-lead (PbPb) collisions
at LHC. In addition to its value in probing several aspects of 
QCD dynamics, photoproduction of heavy quarks can also aid in 
elucidating and constraining some components of nuclear parton 
distributions. Although the parton distribution dependence is
linear and different modifications are superimposed due to the
integration over parton momentum fraction $x$, 
both cross sections and rapidity distributions for $c\bar{c}$ in PbPb
collisions manifest appreciable sensitivity to
shadowing around midrapidity and a slight sensitivity to antishadowing 
at more forward and backward rapidities. Thus $c\bar{c}$
photoproduction offers good constraining potential for shadowing, and
a somewhat less potential for antishadowing. Despite the fact that 
photoproduction of $b\bar{b}$ is less sensitive to nuclear
modifications than $c\bar{c}$, the influence of shadowing is evident 
around midrapidity, and it thus offers some constraining ability for 
shadowing. Both $c\bar{c}$ and $b\bar{b}$ total photoproduction cross 
sections and rapidity distributions in pPb collisions show little sensitivity
to nuclear modifications and are therefore useful in shadowing determination
via cross section ratios. In addition their resolved components are 
appreciable, especially for $b\bar{b}$ and thus it seems feasible that 
they could be of some use in constraining photon parton distributions.  

\bigskip
We thank C.A. Bertulani and M.J. Murray for helpful suggestions.
%

\end{document}